\documentstyle[prd,aps,preprint,tighten]{revtex}
\begin{document}
\title{A simple Dirac wave function for\\ a Coulomb potential
with linear confinement}
\author{Jerrold Franklin\footnote{Internet address:
V5030E@VM.TEMPLE.EDU}}
\address{Department of Physics\\
Temple University, Philadelphia, PA 19122-6082}
\date{December 22, 1998}
\maketitle
\begin{abstract}
A simple analytical solution is found to the Dirac equation for the combination
of a Coulomb potential with a linear confining potential.  An appropriate linear
combination of Lorentz scalar and vector linear potentials, with the scalar part
dominating, can be chosen to give a simple Dirac wave function.  The binding
energy depends only on the Coulomb strength and is not affected by the linear
potential.  The method works for the ground state, or for the lowest state with
$l=j-\frac{1}{2}$, for any $j$.
\end{abstract}
\pacs{PACS number(s): 03.65.Pm, 12.39.Ki, 12.39.Pn}

It is often useful in quark model calculations to have a simple wave function to
help in parameterizing  physical quantities.  A variety of non-relativistic and
relativistic wave functions have been used for this purpose, but the desire for
simplicity has often limited the physical reasonableness of the wave function.
In this paper, we propose a simple relativistic wave function that is the
solution of the Dirac equation for a potential that has the general features
that have been suggested by QCD.  The potential we use has a Coulombic part and
a linear confining part.  The following calculation is exceedingly simple, but
we have not seen it presented anywhere.  We present it here so that it might be
useful to anyone who would like to use a simple, but reasonable, relativistic
potential and wave function.

The Dirac equation we solve is
\begin{equation}
[{\bf\alpha\cdot p}+\beta m -\frac{\lambda}{r} +\beta\mu r + \nu r]\psi=E\psi,
\end{equation}
where {\boldmath$\alpha$\unboldmath} and $\beta$ are the usual Dirac matrices.
The linear confining part has been split into two parts, $\beta\mu r$ which is a
Lorentz scalar potential, and $\nu r$, a Lorentz vector potential.  We will see
that the inequality $\mu > \nu$ is necessary for the potential to lead to a
bound state.  That is, the Lorentz scalar part must dominate over the Lorentz
vector part of the confining potential.  The four component wave function $\psi$
can be written in terms of two component spinors $u$ and $v$ as
\begin{equation}
\psi =  N\left(\begin{array}{c}u\\ v\end{array}\right),
\end{equation}
with $N$ an appropriate normalization constant.
The two component spinors $u$ and $v$ satisfy the equations
\begin{eqnarray}
({\bf\sigma\cdot p})u & = & [E+m+\frac{\lambda}{r}+(\mu-\nu) r]v\\
({\bf\sigma\cdot p})v & = & [E-m+\frac{\lambda}{r}-(\mu+\nu) r]u.
\end{eqnarray}

For $u$, we choose the function
\begin{equation}
u=r^{b-1}e^{-ar}e^{-\frac{1}{2}\alpha^2 r^2}\chi,
\end{equation}
where $\chi$ is a constant two component spinor.
This is just the form of the exact Coulomb solution with an added factor
$e^{-\frac{1}{2}\alpha^2 r^2}$.  Acting on this form of $u$ with the operator
{\boldmath$\sigma\cdot p$\unboldmath} gives the result that
\begin{equation}
({\bf\sigma\cdot p})u  = i({\bf\sigma\cdot{\hat r}})[a+\frac{1-b}{r}+\alpha^2
r]u.
\end{equation}
Comparision of Eqs. (3) and (6) for $({\bf\sigma\cdot p})u$ shows that $v$ can
be written as
\begin{equation}
v=i\gamma({\bf\sigma\cdot{\hat r}})u,
\end{equation}
where $\gamma$ is a constant factor, given by
\begin{equation}
\gamma=\frac{a}{m+E}=\frac{1-b}{\lambda}=\frac{\alpha^2}{\mu - \nu}.
\end{equation}
The three equalities in Eq.\ (8) result from comparing the constant term, the
$\frac{1}{r}$
term, and the $r$ term in Eqs.\ (3) and (6) separately.
The form for $v$ given by Eq.\ (7) can now be substituted into Eq.\ (4).  This
results in the same form for $u$ as in Eq.\ (5), and leads to three new
equations for the constant $\gamma$
\begin{equation}
\gamma=\frac{m-E}{a}=\frac{\lambda}{1+b}=\frac{\mu + \nu}{\alpha^2}.
\end{equation}

The relations in Eqs.\ (8) and (9) can be combined, resulting in
\begin{eqnarray}
b & = & \sqrt{1-\lambda^2}\\
a & = & m\lambda\\
E & = & m\sqrt{1-\lambda^2}.
\end{eqnarray}
These are just the same results, including the same energy, as for a simple
Coulomb potential without the added linear confinement.  The confining potential
constants must be related by
\begin{equation}
\nu=-\mu\sqrt{1-\lambda^2},
\end{equation}
so that the vector part of the linear potential must be less in magnitude than
the scalar part.
 The Gaussian constant $\alpha^2$ is given by
\begin{equation}
\alpha^2=\lambda\mu,
\end{equation}
so that $\mu$ must positive and $\nu$ negative.  Note that the scalar
confining strength can take any value, as long as $\mu$ is positive and
the two constants,
$\mu$ and $\nu$, are related by Eq.\ (13).
We see that the solution is not completely general, but can only work if the
three potential constants satisfy the constraint of Eq.\ (13).  However this
still permits a wide range of confining potentials, because the sum $\mu+\nu$ is
unconstrained.  Only the ground state will have the simple analytic form of
Eqs.\ (5) and (7).   The radially excited states could not be written so simply,
and their energies would depend on the confinement constants $\mu$ and $\nu$ as
well as on $\lambda$ .

The method also works for the lowest excited state for which $l=j-\frac{1}{2}$.
In this case, equations (11)-(14) get modified by the substitution
\begin{equation}
\lambda\rightarrow \lambda/\kappa,
\end{equation}
where $\kappa=j+\frac{1}{2}$ is the principal quantum number of the state.
This can be seen by writing the Dirac wave function in terms of  radial
functions f(r) and g(r)
\begin{equation}
\psi =  N\left(\begin{array}{c}
f(r){\cal Y}^j_{j-\frac{1}{2}}\\
-ig(r){\cal Y}^j_{j+\frac{1}{2}}
\end{array}\right),
\end{equation}
where ${\cal Y}^j_l$ is a two component angular spinor function corresponding to
total angular momentum $j$ and orbital momentum $l$.
The radial functions satisfy the equations
\begin{eqnarray}
\left(\frac{d}{dr}+\frac{1-\kappa}{r}\right)f & = &
-\left[E+m+\frac{\lambda}{r}+(\mu-\nu) r
\right]g\\
\left(\frac{d}{dr}+\frac{1+\kappa}{r}\right)g & = &
\left[E-m+\frac{\lambda}{r}-(\mu+\nu) r
\right]f.
\end{eqnarray}

For the function f, we choose the form
\begin{equation}
f=r^{b-1}e^{-ar}e^{-\frac{1}{2}\alpha r^2}.
\end{equation}
Then substituting this form into Eqs. (17) and (18) leads to the result
\begin{equation}
g=\gamma f
\end{equation}
with
\begin{equation}
\gamma=\frac{\kappa-b}{\lambda}=\frac{\lambda}{\kappa+b}
=\frac{\alpha^2}{\mu-\nu}=\frac{\mu+\nu}{\alpha^2}
=\frac{a}{m+E}=\frac{m-E}{a}.
\end{equation}
Equation (21) leads to the result
\begin{equation}
b=\sqrt{\kappa^2-\lambda^2},
\end{equation}
and the modification $\lambda\rightarrow\lambda/\kappa$ in Eqs. (11)-(14).

In summary, we have presented a simple Dirac wave function for a Coulomb
potential with linear confinement.

\end{document}